# Correlated fragile site expression allows the identification of candidate fragile genes involved in immunity and associated with carcinogenesis

Angela Re*[1], Davide Cora[1,4], Alda Maria Puliti[2,3], Michele Caselle[1,4] and Isabella Sbrana[2]

Address: [1]Dipartimento di Fisica Teorica dell'Università degli Studi di Torino e INFN, Via P. Giuria 1 – I 10125 Torino, Italy, [2]Dipartimento di Biologia dell'Università degli Studi di Pisa, Via San Giuseppe 22 – I 56126 Pisa, Italy, [3]Laboratorio di Genetica Molecolare, Istituto G.Gaslini, L.go Gaslini 5 – I 16147 Genova, Italy and [4]Centro Interdipartimentale Sistemi Complessi in Biologia e Medicina Molecolare, Via Accademia Albertina 13 – I 10123 Torino, Italy

Email: Angela Re* - angelare@to.infn.it; Davide Cora - cora@to.infn.it; Alda Maria Puliti - apuliti@unige.it; Michele Caselle - caselle@to.infn.it; Isabella Sbrana - isbrana@biologia.unipi.it

* Corresponding author

## Abstract

**Background:** Common fragile sites (cfs) are specific regions in the human genome that are particularly prone to genomic instability under conditions of replicative stress. Several investigations support the view that common fragile sites play a role in carcinogenesis. We discuss a genome-wide approach based on graph theory and Gene Ontology vocabulary for the functional characterization of common fragile sites and for the identification of genes that contribute to tumour cell biology.

**Results:** Common fragile sites were assembled in a network based on a simple measure of correlation among common fragile site patterns of expression. By applying robust measurements to capture in quantitative terms the non triviality of the network, we identified several topological features clearly indicating departure from the Erdos-Renyi random graph model. The most important outcome was the presence of an unexpected large connected component far below the percolation threshold. Most of the best characterized common fragile sites belonged to this connected component. By filtering this connected component with Gene Ontology, statistically significant shared functional features were detected. Common fragile sites were found to be enriched for genes associated to the immune response and to mechanisms involved in tumour progression such as extracellular space remodeling and angiogenesis.

Moreover we showed how the internal organization of the graph in communities and even in very simple subgraphs can be a starting point for the identification of new factors of instability at common fragile sites.

**Conclusion:** We developed a computational method addressing the fundamental issue of studying the functional content of common fragile sites. Our analysis integrated two different approaches. First, data on common fragile site expression were analyzed in a complex networks framework. Second, outcomes of the network statistical description served as sources for the functional annotation of genes at common fragile sites by means of the Gene Ontology vocabulary. Our results support the hypothesis that fragile sites serve a function; we propose that fragility is linked to a coordinated regulation of fragile genes expression.

## Background

Fragile sites are hot spots more sensitive to sister chromatid exchange and recombination, plasmid and DNA viral integration and amplicons than other regions of the human genome.

Fragile sites are thought as an inherent component of chromosomal structure. These sites can extend over large DNA sequences, often up to several Mbp in length. They are said to be 'expressed' when they exhibit cytogenetic abnormalities that appear as gaps or breaks on metaphase chromosomes. Fragile site expression can be elicited by treatment of cells with aphidicolin that inhibits DNA polymerases $\alpha$ and $\delta$. At present 84 common fragile sites (75 out of them are aphidicolin-inducible) are listed in the genome database GDB [37]. The exact number of common fragile sites is a matter of interpretation because increasing the stress placed on DNA replication leads to the expression of an increasing number of fragile sites. Fragile site susceptibility to even low doses of replication inhibitors suggests that they are regions intrinsically difficult to replicate [1].

Molecular sequence analyses reveal no intrinsic characteristics of these regions that might explain their instability [2]. However, these unstable regions are evolutionarily conserved; this suggests that they may have a functional role [3]. Up to now sequencing shows the presence of AT-rich regions characterized by high flexibility [4] and low stability but they do not seem sufficient by themselves to explain fragility.

More progress has been made in the identification of factors acting *in trans* to regulate fragile site expression. Indeed fragile site stability is thought to be ensured by chromosome-bound signal transduction proteins that mediate checkpoint responses during cell cycle.

Such a hypothesis is supported by several pieces of evidence. In particular, a few papers demonstrate that ATR activity may protect fragile sites from their expression proposing different models [5-8].

Over the past few years common fragile sites have become an important issue in cancer biology. Indeed the requirement of several mutations for tumorigenesis and the fact that most cancers harbour a large number of genetic and/or epigenetic changes have led to suggestions that fragile site-associated instability is a hallmark of tumorigenesis. Several studies demonstrate that at FRA3B and FRA16D (the most expressed fragile sites) large, intra-locus deletions or translocations often alter genes such as FHIT and WWOX [9]. Both genes show tumour suppressor functions. In addition, fragile sites initiate breakage-fusion-bridge (BFB) cycles responsible for gene amplification. Amplicons, which partially map to FRA7G and involve the MET oncogene, have been recently found in six primary esophageal adenocarcinomas [10]. These findings strongly corroborate the hypothesis that altered genetic expression due to rearrangements at fragile sites could have a causal role in cancer.

However recent reports challenge the prevailing view asking whether, given that common fragile sites are a normal component of human chromosomes, they can play a protective role against cancer at the incipient tumour stages. Experimental evidences show that tumorigenic events activate the ATR/ATM-regulated checkpoint through deregulated DNA replication and DNA damage and thereby activate an inducible barrier against tumour progression [11,12]. Fragile site-associated instability could take part to the cellular counter-response against oncogenic stress.

The above evidences suggest two possible scenarios: fragile sites can promote tumour progression or can act as 'sensors' to elicit, by altered expression of their genes, cellular response against hazards at preliminary stages. One of the aims of our work is to shed some light on this issue by looking at the common functions of genes located at fragile sites which show correlated expression patterns.

Most of the studies on fragile sites since their discovery were based on a site-by-site approach. In this paper we try to reconsider this modeling paradigm and ask whether our understanding of fragile sites might benefit from a comparative description. In our opinion the tools elaborated in the last years in the context of complex networks theory offer the ideal framework to try a description of fragile sites as an interconnected system.

We are motivated by the observation of a correlation between breakage levels at two very frequent fragile sites (FRA3B and FRA16D) and at a number of less frequent other ones in lymphocytes from subjects exposed to radiation and affected by radiation-induced thyroid cancer reported in [13]. The cells from the subjects showing the highest fragility were also characterized by a reduced ability to undergo apoptosis, which is a well-known function of the fragile genes FHIT and WWOX, mapping to FRA3B and FRA16D respectively. These findings suggest that genes located at fragile sites share functions that could be involved in a common biological process and that fragility at fragile sites, by altering genetic expression, could somehow bias such process.

The simplest and most effective measure to assemble fragile sites in a relational network is the correlation of their expression profiles on a controlled sample. We apply robust tools and measurements to capture in quantitative terms the non triviality of the network which should be encoded in its topology.

We then suggest that the theoretical efforts mentioned above serve as a driving source to uncover the functional properties of fragile sites. We think that if the topology of the co-expression network indeed deviates from a random graph it should somehow correspond to a coordination among factors of instability placed at fragile sites. To clarify functional implications of the topological analysis outcomes we filter highly significant substructures (to be defined below) by means of the Gene Ontology (GO) functional annotation scheme [14].

Gene Ontology provides a dynamic, controlled vocabulary for describing gene products in any organism. GO includes three extensive subontologies describing molecular function (the biochemical activity of a gene product), biological process (the biological goal a gene product contributes to) and cellular component (the cellular place where the biological activity of a gene product is exerted). Each term has an accession number, a name, a more detailed definition and other information relating this term to its parent terms. Individual terms are organized as a directed acyclic graph, in which the terms form the nodes in the ontology and the links the relationships. Descendent terms are related to their parent terms by 'is-a' relationships or 'part-of' relationships. In contrast to simpler hierarchical structures, one node in a directed acyclic graph may have multiple parents. This allows for a more flexible and detailed description of biological functions. The GO terms do not themselves describe specific genes or gene products; instead, collaborating databases generate associations of GO terms to specific gene products. Gene products are annotated at the most specific level possible, but are considered to share the attributes of all ancestor terms.

## Results and discussion

### *Graph theoretical analysis allows to identify strictly interrelated fragile sites*

Our starting point is the set of 116 significantly expressed fragile sites selected as discussed in details in the Methods section. Fragile site definition is a matter of interpretation based on criteria for inclusion and statistical analysis of data. To detect non-random breakpoints we adopt the approach described in [15] under the proportional probability (PPM) assumption (the probability of a random break at a region is proportional to the region width). In this paper the authors suggest an iterative procedure that recognizes fragile sites using highest observed breakages.

Interestingly 68 out of our 116 fragile sites selected in this way are annotated as aphidicolin-sensitive common fragile sites at the NCBI database [37].

We then evaluate the Spearman's rank-order correlation coefficient between the expression patterns of each pair of fragile sites (see again the Methods section for definitions) and we select only those pairs with a correlation higher than a given threshold.

More precisely we set three thresholds:

$r_s$ = 0.562. This threshold corresponds to correlators which have only a (Bonferroni corrected) probability of 1% to appear by chance. This choice allows us to pick up only highly correlated pairs of fragile sites.

$r_s$ = 0.527. This choice allows us to select all pairs of fragile sites showing a relevant degree of correlation (only 5% of Bonferroni corrected probability to be selected by chance).

$r_s$ = 0.511. This choice allows us to select all pairs of fragile sites showing a significant correlation (10% of Bonferroni corrected probability to be selected by chance).

Let us examine our findings in these three cases in detail. At each significance level connected components are named by capital letters. If, by reducing the stringency of significance for fragile site correlation, some connected components merge into a new enlarged one then the label of this last one includes multiple letters, one for each merged connected component. In the following discussion of features related to connected components, we always specify the significance level at which we refer in the label of the connected components themselves.

#### *First threshold*

*Highly correlated sites* ($r_s \geq 0.562$). 46 correlation coefficients, out of the possible 6670, survive this selection. The fragile sites linked by these correlators turn out to be organized in a main connected component ($A_1$) including 18 nodes joined by 40 links.

Remaining nodes are organized in an isolated link ($C_1$) and in a set of 6 nodes joined by 5 links ($B_1$). A visualization of the whole network is reported in Figure 1.

Fragile sites, organized as just described, are listed in Tab 1.

#### *Second threshold*

($r_s \geq 0.527$). Above this threshold 63 correlation coefficients survive.

The majority of these correlators (57 out of 63) collapse again in a main connected component ($A_5$) composed by 21 vertices. Besides this large component, the other 2 subgraphs ($B_5$, $C_5$) are still present. A full list of correlated

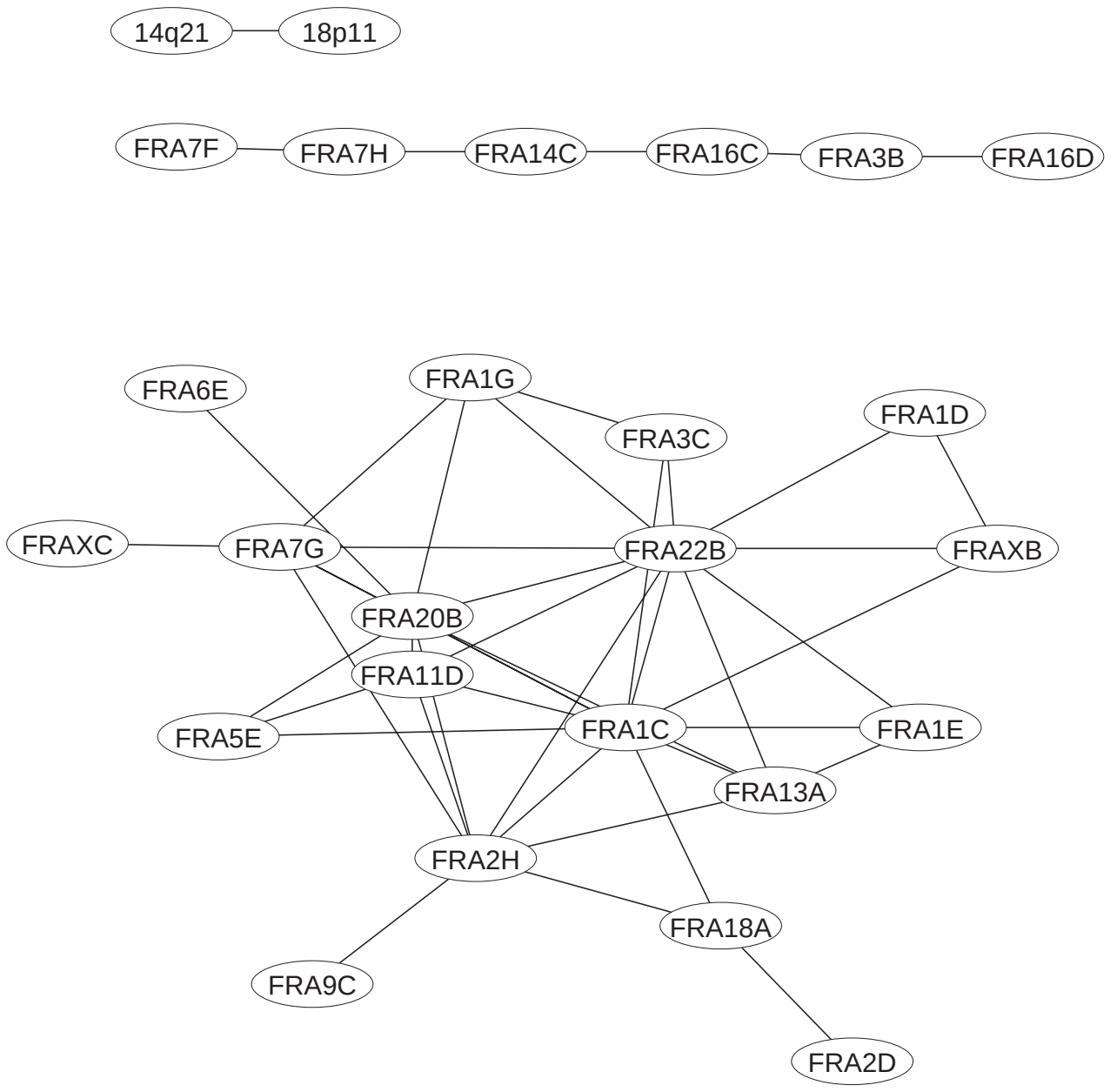

**Figure 1**
Visualization of the network based on correlated expression patterns for fragile sites at $\alpha$ = 1%.

fragile sites at this threshold is given in Tab 1. The network is reported in Figure 2.

*Third threshold*

($r_s \geq 0.511$). At this threshold 84 correlation coefficients exist among 33 fragile sites. The most evident change concerns the main connected component which absorbs both the set made of 6 nodes and the single link mentioned at the upper thresholds ($ABC_{10}$). The main connected component increases its size up to 31 nodes and involves a total of 83 links. Besides this feature, only a single link appears ($D_{10}$). A full list of correlated fragile sites at this threshold can be found again in Tab 1. The network is reported in Figure 3.

It is important to stress that the presence of a large connected component (the one with 18 vertices) at the highest threshold is a highly non trivial and unexpected result. This is best appreciated by comparing our finding with what we would expect for a standard Erdos-Renyi random graph with the same number of links.

The highest threshold discussed above corresponds to a link density p = 0.0069 i.e. to a mean expected degree z = 0.79 which is far below the percolation threshold (which is located at z = 1 for this class of graphs). This observation suggests that the above connected component is definitely non-random and should contain some kind of biological information. We shall deal with the problem of extracting this information in the next section.

In order to evaluate in a more quantitative way the features of the networks that we find and also to measure the importance of the nodes in the network we used three

**Table 1: Clustered fragile sites as they are detected by means of the Hoshen-Kopelman algorithm at the three significance thresholds (see text).**

| Threshold | Cluster | Nodes | Links | Fragile sites | | | | | | | | | |
|---|---|---|---|---|---|---|---|---|---|---|---|---|---|
| 1% | $A_1$ | 18 | 40 | FRA1C | FRA1D | FRA1E | FRA1G | FRA2D | FRA2H | FRA3C | FRA5E | FRA6E | FRA7G |
| | | | | FRA9C | FRA11D | FRA13A | FRA18A | FRA20B | FRA22B | FRAXB | FRAXC | | |
| | $B_1$ | 6 | 5 | FRA3B | FRA7F | FRA7H | FRA14C | FRA16C | FRA16D | | | | |
| | $C_1$ | 2 | 1 | 14q21 | 18p11 | | | | | | | | |
| 5% | $A_5$ | 21 | 57 | FRA1C | FRA1D | FRA1E | FRA1G | FRA2D | FRA2H | FRA3C | FRA5D | FRA5E | FRA6E |
| | | | | 7p14 | FRA7G | FRA9C | FRA10F | FRA11D | FRA13A | FRA18A | FRA20B | FRA22B | FRAXB |
| | | | | FRAXC | | | | | | | | | |
| | $B_5$ | 6 | 5 | FRA3B | FRA7F | FRA7H | FRA14C | FRA16C | FRA16D | | | | |
| | $C_5$ | 2 | 1 | 14q21 | 18p11 | | | | | | | | |
| 10% | $ABC_{10}$ | 31 | 83 | FRA1C | FRA1D | FRA1E | FRA1G | 1q43 | FRA2D | FRA2H | FRA3B | FRA3C | FRA5D |
| | | | | FRA5E | FRA6E | 7p14 | FRA7B | FRA7F | FRA7G | FRA7H | FRA9C | FRA10F | FRA11D |
| | | | | FRA13A | FRA14C | 14q21 | FRA16C | FRA16D | 18p11 | FRA18A | FRA20B | FRA22B | FRAXB |
| | | | | FRAXC | | | | | | | | | |
| | $D_{10}$ | 2 | 1 | FRA5C | 7p15 | | | | | | | | |

Capital letters label connected components at each significance level $\alpha$ that is reported in the left-most column. Labels including multiple letters reflect connected components merging. Numbers in labels indicate the significance level at which connected components are detected. The third and fourth columns include respectively the number of fragile sites and of internal links for each connected component. Fragile sites are reported in the right-most column.

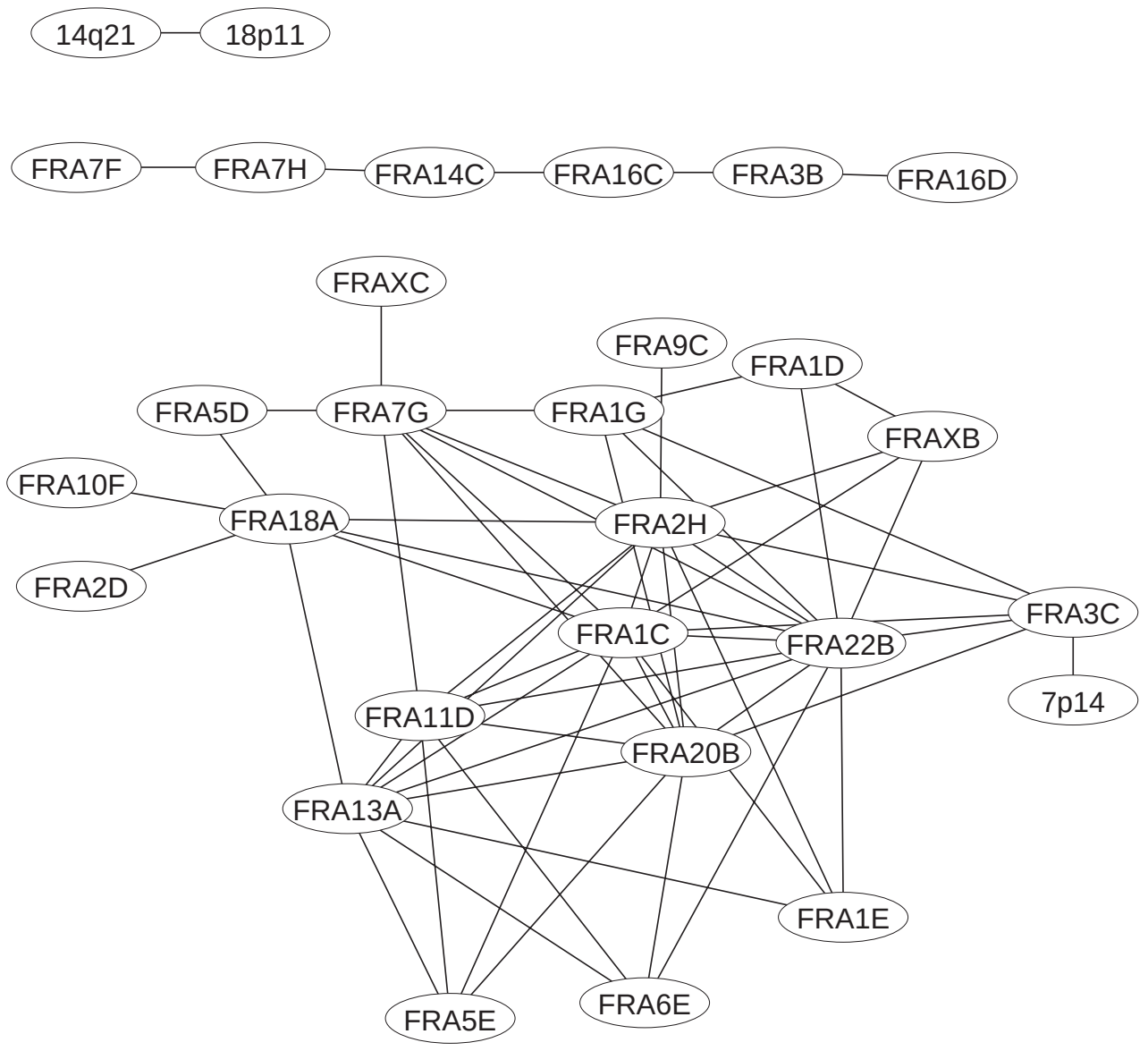

**Figure 2**
Visualization of the network based on correlated expression patterns for fragile sites at $\alpha$ = 5%.

indices which are by now standard tools in the graph theoretical analysis of networks: degree, betweenness and clustering coefficient.

### *Degree*

The simplest measure is *degree*, which is the number of links connected to a vertex in a network. The probability of finding a vertex of degree k in a Erdos-Renyi random graph is given by a Poisson distribution (see Methods section). We see that, looking for instance at the highest threshold network, the probability of having a vertex of degree 2 is 0.14 while (what is more impressive) the probability of having a vertex of degree 11 is 8.50e-10. This is the case for instance of FRA22B. It is interesting to observe that even at lower thresholds the connectivity degree of FRA22B remains remarkably high with respect to the random graph expected probability: at the intermediate threshold the degree of FRA22B becomes 13 (probability 1.48e-10) while at the lowest threshold its connectivity becomes 14 (probability 4.89e-10). Besides FRA22B, the number of fragile sites exhibiting degree higher than 2 in the three networks is in any case very high and increases from 13 at $\alpha$ = 1% to 15 at $\alpha$ = 5% and finally to 19 at $\alpha$ = 10%. Data are available in Tab 2.

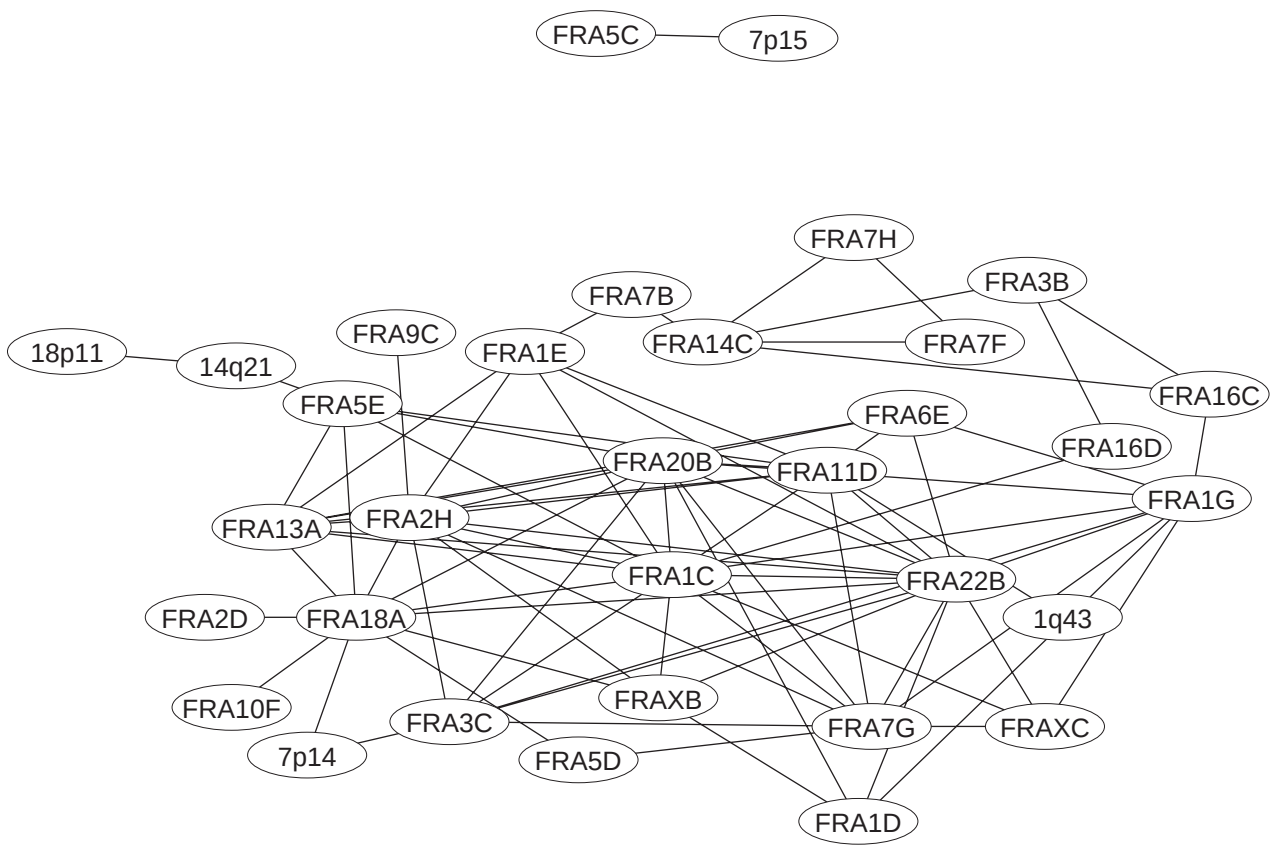

**Figure 3**
Visualization of the network based on correlated expression patterns for fragile sites at $\alpha$ = 10%.

### *Betweenness*

A more sophisticated centrality measure is *betweenness*. Betweenness is a measure of the extent to which a node lies on the paths between others. We define the betweenness of a node *i* as the fraction of shortest paths between pairs of nodes in the network that pass through *i*. Values are reported in Tab 3.

### *Clustering*

An even more useful tool to identify non random features in biological networks is the property of clustering (sometimes also called *network transitivity*). It can be measured using the *clustering coefficient* C. It is essentially the mean probability that two vertices that are network neighbours of the same other vertex are also neighbours. We calculate the clustering coefficient for the whole network. In an Erdos-Renyi random graph C can be easily evaluated and coincides with p (the link density) whose value is very small in all the three graphs. On the contrary in our graphs the clustering coefficient has remarkably high values, with a ratio between the values that we find and the Erdos-Renyi ones higher than 30 (see Tab 4).

This strong tendency of the expressed and correlated fragile sites to cluster among them suggested us to perform a community analysis of the connected components in all three networks.

### *Community analysis*

Roughly speaking communities are groups of vertices within a connected cluster which have a high density of edges within them and a lower density of edges with other communities. There are by now several algorithmic tools which allow to reconstruct the community structure of a given graph (see Methods section for a discussion of one we used); the quality of the community reconstruction is usually given by the so called "modularity coefficient" Q (see Methods section). The rather high values of the clustering coefficient and of the betweenness (which are usually indicators of a potential good community organization) prompted us to perform a community analysis for our networks.

**Table 2: Deviations of fragile sites' degrees in the correlation-based network from the Erdos-Renyi random model.**

| α | Degree | Erdos-Renyi p-value | Fragile sites | | | | | |
|---|---|---|---|---|---|---|---|---|
| 1% | 11 | 8.50e-10 | FRA1C | FRA22B | | | | |
| | 9 | 1.50e-07 | FRA20B | | | | | |
| | 8 | 1.71e-06 | FRA2H | | | | | |
| | 6 | 1.53e-04 | FRA7G | | | | | |
| | 5 | 0.001 | FRA11D | FRA13A | | | | |
| | 4 | 0.007 | FRA1G | | | | | |
| | 3 | 0.037 | FRA1E | FRA3C | FRA5E | FRA18A | FRAXB | |
| | 2 | 0.142 | FRA1D | FRA3B | FRA7H | FRA14C | FRA16C | |
| 5% | 13 | 1.48e-10 | FRA22B | | | | | |
| | 11 | 1.98e-08 | FRA1C | FRA2H | | | | |
| | 10 | 2.02e-07 | FRA20B | | | | | |
| | 9 | 1.87e-06 | FRA13A | | | | | |
| | 8 | 1.56e-05 | FRA7G | FRA11D | | | | |
| | 7 | 1.15e-04 | FRA18A | | | | | |
| | 6 | 7.48e-04 | FRA3C | | | | | |
| | 5 | 0.004 | FRA1G | | | | | |
| | 4 | 0.019 | FRA1E | FRA5E | FRA6E | FRAXB | | |
| | 3 | 0.071 | FRA1D | | | | | |
| | 2 | 0.198 | FRA3B | FRA5D | FRA7H | FRA14C | FRA16C | |
| 10% | 14 | 4.89e-10 | FRA1C | FRA22B | | | | |
| | 12 | 4.23e-08 | FRA20B | | | | | |
| | 11 | 3.50e-07 | FRA2H | FRA18A | | | | |
| | 10 | 2.66e-06 | FRA11D | | | | | |
| | 9 | 1.83e-05 | FRA1G | FRA7G | FRA13A | | | |
| | 7 | 6.27e-04 | FRA3C | | | | | |
| | 6 | 0.003 | FRA1E | FRA5E | | | | |
| | 5 | 0.013 | FRA6E | FRA14C | FRAXB | | | |
| | 4 | 0.043 | FRA1D | FRAXC | | | | |
| | 3 | 0.119 | FRA3B | FRA16C | | | | |
| | 2 | 0.247 | FRA5D | 7p14 | FRA7B | FRA7F | FRA7H | 14q21 |
| | | | FRA16D | | | | | |

Fragile sites with degree higher than 2 are reported at each significance level α. Probabilities of fragile sites having observed degrees in the random model are reported as well.

We find that the network at the lowest threshold can be very clearly divided into two communities which coincide almost exactly with the connected components that we observe at higher levels of the threshold.

In turn these connected components are at this point very well defined and show no evidence of further organization in subcommunities. Indeed they keep their identity even if we enhance the stringency level up to α = 1% (see Tab 5). Remarkably enough this clean separation in communities is also reflected in a sharp separation at the level of GO annotations, a fact which will play a major role in the following discussion. These findings confirm the general impression that the network organization of most common fragile sites is biologically relevant and support the hypothesis that fragile sites serve a function. We shall make use of all these results in the functional analysis of the next section.

#### *Functional characterization of connected components by Gene Ontology tool*

Once equipped with the described network of fragile sites, our further goal is to find out functional relationships among sites forming the network, which up to now have been thought to be functionally independent.

At each significance level the network based on correlated expression patterns is characterized by a non trivial inner organization. Two stable connected components are detected at α = 1% and α = 5%. At α = 10% they are still recognizable as different communities albeit they merge in a unique connected component.

Our aim at this point is to understand if this network structure observed among fragile sites implies some kind of functional interaction among the genes which are contained in such fragile sites.

**Table 3: Fragile site's betweenness and degree measures at each significance level α.**

| Fragile site | Betweenness | Degree | Betweenness | Degree | Betweenness | Degree |
|---|---|---|---|---|---|---|
| | α = 1% | | α = 5% | | α = 10% | |
| FRA1C | 0.121 | 11 | 0.055 | 11 | 0.170 | 14 |
| FRA22B | 0.103 | 11 | 0.104 | 13 | 0.076 | 14 |
| FRA20B | 0.078 | 9 | 0.030 | 10 | 0.060 | 12 |
| FRA2H | 0.086 | 8 | 0.089 | 11 | 0.082 | 11 |
| FRA7G | 0.058 | 6 | 0.077 | 8 | 0.038 | 9 |
| FRA11D | 0.004 | 5 | 0.011 | 8 | 0.080 | 10 |
| FRA13A | 0.004 | 5 | 0.025 | 9 | 0.028 | 9 |
| FRA1G | 0.004 | 4 | 0.011 | 5 | 0.135 | 9 |
| FRA1E | 0.000 | 3 | 0.000 | 4 | 0.103 | 6 |
| FRA3C | 0.002 | 3 | 0.052 | 6 | 0.028 | 7 |
| FRA5E | 0.000 | 3 | 0.000 | 4 | 0.115 | 6 |
| FRA18A | 0.053 | 3 | 0.114 | 7 | 0.170 | 11 |
| FRAXB | 0.005 | 3 | 0.004 | 4 | 0.005 | 5 |
| FRA1D | 0.000 | 2 | 0.001 | 3 | 0.003 | 4 |
| FRA3B | 0.013 | 2 | 0.011 | 2 | 0.017 | 3 |
| FRA7H | 0.013 | 2 | 0.011 | 2 | 0.000 | 2 |
| FRA14C | 0.020 | 2 | 0.016 | 2 | 0.120 | 5 |
| FRA16C | 0.020 | 2 | 0.016 | 2 | 0.088 | 3 |
| FRA6E | _ | _ | 0.000 | 4 | 0.003 | 5 |
| FRA5D | _ | _ | 0.004 | 2 | 0.001 | 2 |
| FRAXC | _ | _ | _ | _ | 0.000 | 4 |
| 7p14 | _ | _ | _ | _ | 0.001 | 2 |
| FRA7B | _ | _ | _ | _ | 0.064 | 2 |
| FRA7F | _ | _ | _ | _ | 0.000 | 2 |
| 14q21 | _ | _ | _ | _ | 0.058 | 2 |
| FRA16D | _ | _ | _ | _ | 0.043 | 2 |

To this end we decided to explore the functional and biological content of the network by means of the Gene Ontology annotation scheme.

We proceed in two steps.

First, we consider the previously outlined connected components and we explicitly list all the genes included in them. At this point, each connected component is associated with a set of genes, collected together only by virtue of their membership to fragile sites with similar expression profile (see Methods for details).

Second, we analyse each of these sets of genes separately, looking for shared biological functions, according the Gene Ontology vocabulary (see Methods section for details) to identify genes that are more likely to be functionally related than expected by chance.

**Table 4: Comparison between clustering coefficients for the whole real network and for the Erdos-Renyi model at each significance level α.**

| α | C | Erdos-Renyi estimate |
|---|---|---|
| 1% | 0.349 | 0.007 |
| 5% | 0.327 | 0.009 |
| 10% | 0.380 | 0.013 |

To this end, a Gene Ontology filter is applied to each connected component of the network. By filtering the content of genes at correlated fragile sites we find that clustered fragile sites are overall enriched in fourteen GO terms. For the discussion herein we select GO terms which turned out stably enriched at each significance level, after correction for multiple test, and are not biased by positional constraints of genes annotated to such terms (see detailed discussion in the Methods section). The eight GO terms obtained in this way are: cytokine activity, hematopoietin/interferon-class (D200-domain) cytokine receptor, interferon-alpha/beta receptor binding, response to virus, extracellular space, carboxylesterase activity, serine esterase activity and xenobiotic metabolism.

Finally, we manually integrate the knowledge get by Gene Ontology with information made available from the Kyoto Encyclopedia of Genes and Genomes (KEGG) pathways database website [39] and the GeneCards database website [40].

**Table 5: Comparison between community organization at $\alpha$ = 10% and connected component organization at $\alpha$ = 5%.**

| | | | | | | | | | | | | | |
|---|---|---|---|---|---|---|---|---|---|---|---|---|---|
| **cluster organization at $\alpha$ = 5% :** | | | | | | | | | | | | | |
| $A_5$ | FRA1C | FRA1D | FRA1E | FRA1G | FRA2D | FRA2H | FRA3C | FRA5D | FRA5E | FRA6E | 7p14 | FRA7G | FRA9C |
| | FRA10F | FRA11D | FRA13A | FRA18A | FRA20B | FRA22B | FRAXB | FRAXC | | | | | |
| $B_5$ | FRA3B | FRA7F | FRA7H | FRA14C | FRA16C | FRA16D | | | | | | | |
| $C_5$ | 14q21 | 18p11 | | | | | | | | | | | |
| **community organization at $\alpha$ = 10% :** | | | | | | | | | | | | | |
| com. no 1 | FRA1C | FRA1D | FRA1E | FRA1G | 1q43 | FRA2D | FRA2H | FRA3C | FRA5D | FRA5E | FRA6E | 7p14 | FRA7B |
| | FRA7G | FRA9C | FRA10F | FRA11D | FRA13A | *14q21* | *18p11* | FRA18A | FRA20B | FRA22B | FRAXB | FRAXC | |
| com. no 2 | FRA3B | FRA7F | FRA7H | FRA14C | FRA16C | FRA16D | | | | | | | |

Communities identified at $\alpha$ = 10% appear as connected components at more stringent significance levels. The two communities at $\alpha$ = 10% coincide respectively with $A_5$ and $B_5$ at $\alpha$ = 5% except for $C_5$ which is enclosed in the larger community. $C_5$'s fragile sites are highlighted by italic font typing.

### *GO terms related to immune response are significantly overrepresented in fragile sites*

We perform the GO analysis for each connected component at the three different significance levels. GO terms that turned out to be overrepresented for at least one significance level $\alpha$ are listed in Tab 6. The full sets of genes annotated to each GO term are provided in the supplementary materials [see Additional files 1, 2, 3, 4, 5]. The following analysis refers to the $\alpha$ = 1% case unless otherwise specified. We examine GO terms separately.

The GO term 'cytokine activity' represents one of the most comprehensive GO functions including 21 genes which grow up to 30 at $\alpha$ = 10%. Notably at the lowest significance level genes annotated to 'cytokine activity' are localized on fragile sites of A1 and B1 as well. For example 16q22.1 (band of FRA16C) harbours a cluster of chemokine-like factor genes including CKLF, CKLF2, CKLF3 and CKLFSF4; erythropoietin EPO, located at 7q22 (FRA7F), is a member of the EPO/TPO family and encodes a secreted, glycosylated cytokine provided with antiapoptotic functions in several tissue types; FAM3D, a member of a cytokine-like gene family [16], maps to 3p14.2 (FRA3B). This observation confirms that the functionality described by the GO term 'cytokine activity' is a common feature of fragile sites of both connected components.

A number of genes, which share strictly related functions and belong to the same protein family, are organized in the genome in very close proximity. Fifteen interferon-$\alpha$, $\beta$,$\varepsilon$ family's genes, producing antiviral and antiproliferative responses in cells, map to a narrow locus spanning 364kb at 9p21.3 (FRA9C). Leukaemia inhibitor factor (LIF), which is a cytokine that induces macrophage differentiation and chemotaxis in immune cells, and oncostatin-M (OSM), cytokine encoding a growth regulator which inhibits the proliferation of a number of tumour cell lines, belong to the interleukin-6 family and are 16kb apart at 22q12.2 (FRA22B).

Other genes map to independent loci. For example thrombopoietin (THPO) at 3q27 (FRA3C), ligand for the product of myeloproliferative leukemia virus oncogene MPL, is a megakaryopoietic regulator. The FKBP-associated protein encoded by GLMN, mapped to 1p22 (FRA1D), has a role in the control of the immune response and an antigrowth function and interacts with the proto-oncogene MET [17].

This set of genes is significantly annotated also to two descendants of the upper-level term 'cytokine activity' in the GO graph: '*hematopoietin/interferon class-cytokine receptor binding*' (17 annotated genes) and '*interferon-alpha/beta receptor binding*' (8 annotated genes, partially coinciding with the previous ones). Similarly the full interferon-$\alpha$/$\beta$ family's genes are responsible for GO process '*response to virus*' being above the threshold.

The subontology 'cellular component' refers to the place where a gene product is active. Therefore it is not a surprise the overrepresentation, albeit at a less significant level, of the GO term '*extracellular space*'. Out of 31 annotated genes for this GO term, 21 are annotated to the GO term 'cytokine activity'. Ten genes specifically annotated to the term '*extracellular space*' map to eight fragile sites. The products of genes related to '*extracellular space*' mainly act in processes such as cell-matrix interaction, degradation of matrix components, growth regulation and apoptosis. They are metalloprotesases (such as PLA2G3 at 22q12.2 and CLCA3 at 1p22) and serine proteases (such as PLG at 6q26) and protease activators.

Let us mention a few GO terms that specifically characterize the connected component $B_1$: '*xenobiotic metabolism*' and '*serine esterase activity*'.

'*Xenobiotic metabolism*' seems to fit well with the notion that fragile sites are sensitive to cancerogenic agents and that they are involved in cancers induced by the exposition to these agents. Indeed, the cytochrome P450 (CYP) genes at FRA7F are known to be involved in the metabolism of exogenous cancerogenic agents in cancer tissues [18]. As for the GO function '*serine esterase activity*', let us stress that the expression of a serine esterase gene is

**Table 6: Functional characterization of fragile sites' connected components.**

| GO id | GO term description | Cluster | $-Log_{10}$ (p value) | | |
|---|---|---|---|---|---|
| | | | $\alpha$ = 1% | $\alpha$ = 5% | $\alpha$ = 10% |
| GO:0005126 | hematopoietin/interferon-class | $A_1$, $A_5$, $ABC_{10}$ | 15.090 | 15.408 | 11.807 |
| | (D200-domain) cytokine receptor | $A_1$, $A_5$, $ABC_{10}$ | | | |
| GO:0005132 | interferon-alpha/beta receptor binding | $A_1$, $A_5$, $ABC_{10}$ | 12.394 | 11.902 | 9.704 |
| GO:0009615 | response to virus | $A_1$, $A_5$, $ABC_{10}$ | 11.857 | 10.889 | 6.811 |
| GO:0005125 | cytokine activity | $A_1$, $A_5$, $ABC_{10}$ | 6.252 | 5.843 | 4.919 |
| GO:0005615 | extracellular space | $A_1$, $A_5$ | 5.390 | 4.627 | < threshold |
| GO:0042272 | nuclear RNA export factor complex | $A_1$, $A_5$ | 4.713 | 4.538 | < threshold |
| GO:0005515 | protein binding | $A_1$ | 4.698 | < threshold | < threshold |
| GO:0008320 | protein carrier activity | $ABC_{10}$ | < threshold | < threshold | 4.645 |
| GO:0004091 | carboxylesterase activity | $B_1$, $B_5$ | 7.133 | 7.133 | _ |
| GO:0004759 | serine esterase activity | $B_1$, $B_5$ | 7.133 | 7.133 | _ |
| GO:0006805 | xenobiotic metabolism | $B_1$, $B_5$ | 4.038 | 4.038 | _ |
| GO:0003700 | transcription factor activity | $D_{10}$ | _ | _ | 5.853 |
| GO:0030528 | transcription regulator activity | $D_{10}$ | _ | _ | 4.758 |
| GO:0003676 | nucleic acid binding | $D_{10}$ | _ | _ | 4.149 |

Overrepresented GO terms are reported along with their Gene Ontology identifiers, full descriptions, enriched connected component(s) for those GO terms and negative logarithm of their hypergeometric p-values.

induced in activated T lymphocytes [19]. Interestingly genes mapped to fragile sites of both connected components turn out to be involved in environmental information processing, albeit in distinct forms.

A detailed description of the specific pathways in which genes annotated to significant GO terms are engaged is provided in Tab 7.

To improve our knowledge of the relationship between fragile sites and gene function, we examined the literature looking for genes belonging to the correlated fragile sites that are involved in some of the functions reported above, even if not yet annotated for them. We found that very recently the FHIT gene, containing the hot spot fragility regions of FRA3B, has been shown to be involved in inflammatory response; it directly inhibits the activation of Prostaglandin E2, a key agent in inflammation, and by this way it suppresses cancer cells proliferation [20]. At FRA16D the WWOX gene has an essential role in the cellular susceptibility to TNF-mediated apoptosis [21]. In the same site PLCG2 gene maps, encoding a phospholipase C that is a crucial enzyme in transmembrane signaling, involved in the activation of a variety of growth factor receptors and immune system receptors.

Together these findings show that many genes sharing the membership to fragile sites with correlated expression participate to common pathways, even if often with a poorly understood role; this supports the hypothesis that fragile sites expression is linked to a coordinated regulation of expression of the fragile site associated genes.

To further characterize the functional meaning of the genes identified by GO analysis, we analyse their possible involvement in cancer. Tab 8 reports the full list of such genes along with references to the sources from which we gather information: the GeneCards database, the Atlas of Genetics and Cytogenetics in Oncology and Haematology database and recent papers available on PubMed database. They are involved prevalently in haematological tumours, such as leukemias, but also in breast, lung, colon, prostate carcinomas and others. Thus, having a role in cancer is a largely shared feature for these genes. Most of them (such as PLG, EPO and PBEF1) exhibit broad expression in the remodelling of the tissues surrounding tumour cells and in the metastasis formation. For a number of genes annotated to GO terms a co-localisation with approved cancer genes occurs; this is the case for FAM3D, located 900 kb distal from FHIT gene, with no intervening genes, included by FRA3B and strictly linked to TU3A (protein DRR1- Down-regulated in renal cell carcinoma 1). Moreover INFE1 is located 350 kb distal from the MTAP, CDKN2A and CDKN2B tumour genes, still with no intervening genes, CKLF and CKLF1-4 are located in a cluster strictly linked to CDH5 (VE-cadherin tumour gene). In this respect it is interesting to note that the aforementioned genes are not evolutionary related paralogs. Therefore the regional proximity seems to have a prevalent functional origin due to the need of a coordinated gene expression.

### *Functional information is embedded in network topology*

Since the co-expression network exhibits a large degree of clustering the most natural examples of subgraphs to

**Table 7: KEGG pathways for genes annotated to overrepresented fragile sites.**

**Signal transduction:**

| MAPK signaling pathway | | Notch signaling pathway | | Wnt signaling pathway | | | Calcium signaling pathway | TGF-beta signaling pathway |
| --- | --- | --- | --- | --- | --- | --- | --- | --- |
| PLA2G3 | MAP3K13 | JAG1 | DVL3 | SENP2 | TCF7 | DVL3 | PDE1A | SMAD5 |

Jak-STAT signaling pathway:

| | | | | | | | | | | | | | | |
| --- | --- | --- | --- | --- | --- | --- | --- | --- | --- | --- | --- | --- | --- | --- |
| IFNA1 | IFNA2 | IFNA4 | IFNA5 | IFNA6 | IFNA8 | IFNA10 | IFNA14 | IFNA17 | IFNA21 | IFNB1 | IFNK | IFNW1 | LIF | OSM |
| EPO | L12RB2 | | | | | | | | | | | | | |

**Signaling molecules and interactions:**

| ECM-receptor interaction | | | Cell adhesion molecules (CAMs) | | | Neuroactive ligand-receptor interaction |
| --- | --- | --- | --- | --- | --- | --- |
| LAMC2 | LAMC1 | ITGAV | ITGAV | VCAM1 | NEGR1 | SST |

Cytokine-cytokine receptor interaction:

| | | | | | | | | | | | | | | |
| --- | --- | --- | --- | --- | --- | --- | --- | --- | --- | --- | --- | --- | --- | --- |
| IFNA1 | IFNA2 | IFNA4 | IFNA5 | IFNA6 | IFNA8 | IFNA10 | IFNA14 | IFNA17 | IFNA21 | IFNB1 | IFNK | IFNW1 | LIF | OSM |
| EPO | TNFSF4 | TNFSF18 | L12RB2 | MET | | | | | | | | | | |

**Immune system**:

| Natural killer cell-mediated cytotoxicity | | | | | | Leukocyte transendothelial migration | | Complement-coagulation cascades |
| --- | --- | --- | --- | --- | --- | --- | --- | --- |
| IFNA1 | IFNA2 | IFNA4 | IFNA5 | IFNA6 | IFNA8 | NCF2 | VCAM1 | PLG SERPINC1 SERPINE1 KNG1 |

| Toll-like receptor signaling pathway | | | | | | | | | | | T/B cell receptor signaling pathway |
| --- | --- | --- | --- | --- | --- | --- | --- | --- | --- | --- | --- |
| IFNA1 | IFNA2 | IFNA4 | IFNA5 | IFNA6 | IFNA8 | IFNA10 | IFNA14 | IFNA17 | IFNA21 | IFNB1 | BCL10 |

**Cell motility and comunication**:

| Focal adhesion | | | | | Adherens junction | Tight junctin | Regulation of actin cytoskeleton | | |
| --- | --- | --- | --- | --- | --- | --- | --- | --- | --- |
| TNR | LAMC2 | MET | LAMC1 | ITGAV | MET | INADL | ITGAV | NCKAP1 | LIMK2 |

**Cell growth and death**:

| cell cycle | | Apoptosis |
| --- | --- | --- |
| CDC7 | CDKN2A | FASLG |

**Genetic information processing**:

| SNARE interactions in vesicular transport | | Basal transcription factors | | Ribosome | RNA polymerase | Proteasom e |
| --- | --- | --- | --- | --- | --- | --- |
| SNAP25 | STX6 | TAF1L | GTF2B | RPL5 | POLR2H | NBEA |

| **Lipid metabolism**: | | | **Nucleotide metabolism**: | | | **Xenobiotic metabolism**: | **Glycan metabolim**: |
| --- | --- | --- | --- | --- | --- | --- | --- |
| Glycerophospholipid metabolism | | | Purine | | Pyramidin e | by cytochrome P450 | Glycosphingolipid metabolism |
| DBT | PLA2G3 | ACHE | PDE1A | POLR2H | POLR2H | CYP3A43 | SMPD3 |

GO annotated genes are associated with their specific KEGG pathways. KEGG pathways are assembled based on the pathways' map that the KEGG database itself provides.

**Table 8: Candidate fragile genes in tumours.**

| Hugo ids | Gene description | Fra. site | Involvement in cancer | |
|---|---|---|---|---|
| | | | Notes | References |
| GLMN | glomulin, FKBP associated protein | FRA1D | familial glomangioma | GCd |
| CLCA3 | chloride channel, calcium activated | FRA1D | strictly linked to CLCA2, tumor suppressor gene for breast cancer | Oncogene 2004 (19): 1471-80 |
| EVI-5 | EVI-5 homolog | FRA1D | | Cell 2006 Jan 27; 124(2): 367-80 |
| TNFSF4 | TNF(ligand) superfamily, member 4 | FRA1G | leukemia T-cell, neoplasms | AGCOH, GCd |
| TNFSF18 | TNF(ligand) superfamily, member 18 | FRA1G | neoplasms | AGCOH |
| TNR | tenascin R | FRA1G | neoplasms | AGCOH |
| ANGPTL1 | Angiopoietin-related protein 1 precursor | FRA1G | neoplasms | Cancer Cell 2004 Nov; 6(5):507-16 |
| ITGAV | CD51 antigen | FRA2H | neoplasm metastasis, colon, breast and pancreatic carcinoma | Cancer Metastasis Rev 2003 Mar;22(1):103-15 |
| FAM3D | family with sequence similarity 3, member D | FRA3B | located 900 kb, with no intervening genes, distal from FHIT gene | |
| THPO | thrombopoietin | FRA3C | myeloproliferative disorders | AGCOH, GCd |
| ADIPOQ | adiponectin, C1Q and collagen domain containing | FRA3C | myeloblastic leukemia | Br J Cancer 2006 Jan 16; 94(1):156-60 |
| AHSG | alpha-2-HS-glycoprotein | FRA3C | liver carcinoma | GCd |
| BCL6 | B-cell CLL/lymphoma 6 | FRA3C | lymphoma b-cell, lymphoma non-hodgkins, burkitt lymphoma | Oncogene 2006 Jul 3 |
| PLG | plasminogen precursor | FRA6E | neoplasm metastasis, neoplasms vascular tissue | J Cell Biochem 2005 Oct; 96(2):242-61 |
| HOXA5 | homeo box A5 | 7p15.2 | breast carcinoma | J Biol Chem 2000 Aug ; 275(34):26551-5 |
| HOXA7,9 | homeo box A7, A9 | 7p15.2 | myeloid leukemia | Leuk Res 2000 Oct; 24(10): 849-55 |
| EPO | erythropoietin | FRA7F | myelodyplastic syndromes | AGCOH, GCd |
| AZGP1 | alpha-2-glycoprotein 1, zinc | FRA7F | cancer of breast, prostate (GCd) | AGCOH |
| PBEF1 | pre-B-cell colony enhancing factor 1 | FRA7F | myeloid leukemia | Leukemia 2005 Jun; 19(6): 998-1004 |
| CAV-1,-2 | caveolin 1, caveolin 2 | FRA7G | carcinoma of colon, adenocarcinoma, ovarian epithelial carcinoma | Genomics 2003 Feb; 81(2):105-7 |
| MET | HGF receptor | FRA7G | gastric carcinoma, neoplasm metastasis, carcinoma renal cell | Oncogene 2006 Jan 19; 25(3):409-18 |
| INFB1 | interferon beta 1 | FRA9C | neoplasms, melanoma | AGCOH, GCd |
| IFNE1 | interferon epsilon 1 | FRA9C | located 350 kb distal from MTAP, CDKN2A and CGKN2B tumor genes | |
| IFNA1 | interpheron alpha 1 | FRA9C | leukemia | AGCOH, GCd |
| IFNA2 | interpheron alpha 2 | FRA9C | myeloid leukemia, renal cell cancer | GCd, TGD |
| SMPD3 | sphingomyelin phosphodiesterase 3 | FRA16C | strictly linked to CDH3 (P-cadherin tumour gene) | |
| CKLF | chemokine-like factor | FRA16C | strictly linked to CDH5 (P-cadherin tumour gene) | |
| CES2 | Carboxylesterase 2 precursor | FRA16C | malignant neoplasms, carcinoma of colon | Neoplasia 2005 Apr;7(4): 407-16 |
| NQO1 | NAD(P)H dehydrogenase, quinone 1 | FRA16C | carcinoma of colon, malignant neoplasm of lung, cancer of bladder | Clin Cancer Res 2005 Dec; 11(24):8866-71 |
| LIF | leukemia inhibitory factor | FRA22B | leukemia, myeloma, neuroepitelioma, stricltly linked to HORMAD2 | AGCOH, TGD |
| OSM | oncostatin M | FRA22B | leukemia, myeloma, kaposi sarcoma, melanoma, plasmacitoma | AGCOH, TGD |
| TCN2 | transcobalamin II, macrocytic anemia | FRA22B | colon adenocarcinoma | GCd |
| SEC14L2 | SEC14-like 2 | FRA22B | neoplasms | Cancer Res 2005 Nov 1; 65(21):9807-16 |

We report genes localized at the observed fragile sites which are assessed to play a role in tumour development in public databases or PubMed articles. Abbreviations for cited databases: AGCOH, Atlas of Genetics and Cytogenetics in Oncology and Haematology; TGD, Tumor Gene Database; GCd, Genecards database.

study are triangles. Even though triangles are the simplest substructures, they clearly indicate departures from randomness and should point out future attempts to understand instability at fragile sites.

Here we confine ourselves to showing, for example, that triangles, which are the tightest ways to connect three nodes, seem to reflect acquired knowledge. We choose, to this end, a triangle which appears at the highest threshold and is significantly annotated to all the GO terms found overrepresented in the connected component $A_1$. It includes FRA7G, FRA3C and FRA22B. Interestingly two candidate tumour suppressor genes CAV-1 and CAV-2 localize at FRA7G. They modulate cell cycle progression through several cascades such as the inhibition of the phosphatidylinositol 3-kinase (PI3-kinase)/Akt survival pathway and the control of the Ras-p42/44 MAP kinase cascade.

FRA22B harbours a family of lipid-binding proteins including SEC14L2,-3,-4 which are homolog to the tocopherol-associated protein (TAP). Note that TAP exerts its tumour suppressor function via down-regulation of PI3-kinase/Akt signaling.

The full list of triangles at $\alpha = 1\%$ is reported in the supplementary materials [see Additional file 6].

## Conclusion

Common fragile sites appear as gaps and breaks at non-random loci on metaphase chromosomes. They are generally stable in somatic cells and inducible in cells cultured under conditions of replicative stress. Several investigations have pointed out a direct correlation between chromosomal fragility and DNA instability in various cancer cases. Common fragile sites are hierarchical in their relative frequency of expression with FRA3B being most frequently observed.

Correlated patterns of expression among a few fragile sites have been recently observed by analysing a distinctive sample of subjects, exposed to radiation and affected by cancer reported in [13]. We are thereupon prompted to clarify whether correlated breakage should be a generally shared feature of fragile site expression. In our opinion complex networks formalism provides the ideal tool to answer this question. Fragile sites are described as nodes of a network where nodes are joined in pairs if the correlation in their patterns of expression is statistically significant. We focus on three robust measures of a network's topology: clustering coefficient, division in connected components and communities. In this respect, the most surprising result is the existence of a large connected component at the highest threshold along with the conservation of a well-defined sub-structure organization at each of the significance levels set in the analysis. The computational method we propose supports aforementioned results and extends their validity. Indeed we show that correlated expression involves a growing number of fragile sites (depending on the significance of correlation) independently from carcinogenic exposure. This general feature of fragile sites supports the hypothesis that they serve a function. Assuming that network topology should reflect underlying cellular mechanisms, we explore the functional content of highly interconnected fragile sites by means of the Gene Ontology vocabulary.

The functional characterization through GO of genes located at connected fragile sites clearly highlights that a great proportion of genes with significant annotated terms are involved in innate and adaptive immune responses and in particular in pathways characteristic of activated T lymphocytes. This is of special interest, since the expressed fragile sites in the present work have been detected in activated lymphocytes.

From these findings we propose that correlated breakage at fragile sites may originate in proliferating lymphocytes from a co-regulated modified expression of fragile genes; in this view the genes identified by GO analysis may be new fragile genes; chromatin changes and DNA replication alteration at or near these genes would be produced by cellular processes connected with their co-regulation performed through still unknown mechanisms. This is supported by the observation that a number of the analysed cytokine-related genes show actual functional interactions in lymphocytes or other cell types. For example OSM and LIF, that have common biologic activities, are stimulated by CSF2.

Interestingly, an emerging view on immune response is that programmed gene expression is achieved by means of epigenetic mechanisms tied to the structure of chromatin; in particular epigenetic effects organize the ability of signal transduction pathways to generate a set of functionally characterized cell progeny [22].

Epigenetic changes at fragile genes associated with highly expressed fragile sites, such as FHIT and WWOX, have been characterized in a number of tumours [23] and suggested to be associated with expression of fragility [24]. Thus, epigenetic mechanisms responsible of coordinated regulation of different loci may be involved in the chromatin alterations leading to chromosome fragility; these alterations could become permanent in cells that undertake a tumorigenic process. A surprising high proportion of the genes identified at correlated fragile sites are implicated in cancer. This finding agrees with the ancient hypothesis of a general relationship between fragile sites and cancer. It is supported by the detection at the charac-terized fragile sites of accepted or supposed tumour suppressor genes such as the proapoptotic genes FHIT at FRA3B and Wwox at FRA16D and others.

According to a recent proposal breakage at fragile sites may be protective against cancer [25,11]. Such a protective role would be mediated by breaks formed in consequence of aberrant replication at fragile sequences, known to be difficult to replicate; breaks would represent a signature of replication stress and would activate the DNA damage checkpoints leading to cell-cycle arrest or apoptosis to ensure genomic integrity. This proposal is supported by the evidence that DNA damage response is activated early in the tumorigenesis and that in this phase loss of heterozigosity occurs preferentially at fragile sites, as possible consequence of stalled replication forks [12].

On the basis of our results on fragile site network we propose to extend it by including that replication stress at fragile sequences is coupled with a modified expression of the associated fragile genes. More precisely, we believe that fragile sequences, sensitive to replication stress, are not located by chance within or near fragile genes, but participate together with genes to the mechanism that regulate the cellular response to DNA damage. This proposal fits for a number of known genes mapping at highly expressed fragile sites, such as FHIT at FRA3B, Wwox at FRA16D, CAV1 and CAV2 at FRA7G and others, that have a function in cell proliferation control and apoptosis. We may ask to what extent such a proposal fits also with the genes identified in our analysis, prevalently related to the immune response. In particular could these genes have the above described role in preventing genome instability? Could their response be significant for non-lymphocyte cell types that are also enabled to express chromosome fragility? Similarly to the known fragile genes, for a number of genes here identified a function in cell cycle control has been described.

However also other genes participating specifically in immune response may be involved in cancer related processes; indeed a link between the immune response and processes that regulate genome integrity has been very recently suggested by the evidence that genotoxic stress and stalled DNA replication up-regulate some stimulatory receptors of the innate immune system such as NKG2D receptor [26]. Thus, it has been suggested that DNA damage response, besides arresting cell cycle, enhancing DNA repair or triggering apoptosis may participate in alerting the immune system to the presence of potentially dangerous cells [27]. Of great interest, according to a very recent study, FHIT gene is involved in inflammatory response by inhibiting synthesis of Prostaglandin E2, a key agent in inflammation [20]; this finding clearly defines a function in immunity for the major fragile gene and thus strongly supports our hypothesis that regulation mechanisms of fragile genes expression could be implied in fragility.

This complex relationship needs to be tested experimentally. Nevertheless the candidate fragile genes identified in our study could be investigated as actors in DNA damage response, associated with carcinogenesis and involved in loss of function in primary steps of tumour development.

## Methods

### *Cytogenetic analysis*

Data on breakage events at aphidicolin-sensitive fragile sites have been obtained in three independent analyses carried out on peripheral blood lymphocytes to compare fragile sites expression in unexposed subjects, healthy subjects and in subjects exposed to environmental carcinogens, such as radiations and pesticides [28,29,13]. All analyses have been carried out by using identical cell culture procedures; chromosome breakage was detected by two experienced cytogenetists sharing appointed criteria; this allows reach results with high reproducibility, verified by repeated samplings.

Chromosomes were stained with the standard GTG banding technique (400 band resolution). Band localization was assigned according to the Mitelman Database of Chromosome Aberrations in Cancer ISCN (1995) [30]. For each subject 100 metaphases are scored for gaps, breaks and rearrangements on coded slides; for subjects exposed to radiation (showing very high breakage) 50 metaphases are analysed.

Our original dataset consists of the expression profiles of 343 chromosomal bands measured on a sample of 60 subjects. To test the nonrandomness of breakage at a given chromosomal band we adopt the algorithm described in [15] under the proportional probability assumption. This model assumes that the probability of a random break at a region is proportional to the range width. Basically, to determine whether a chromosomal region is a fragile site or not, an iterative procedure tests the region with the highest observed standardized breakage number. If such a region is accepted as a fragile site then the procedure goes on to the next iteration leaving out this region. The algorithm stops when it gets a subset of regions for which the test is not able to reject the hypothesis of random breakage.

After such analysis, we end up with a dataset of 116 chromosomal bands, so that the raw data on fragile site expression can be embedded in a matrix $M [116] [60]$ whose $m_{ij}$ element represents the absolute number of breakage events (i.e. total gaps, beaks or rearrrangements) that affect the fragile site $i$ in the subject $j$. We provide this matrix in the supplementary materials [see Additional file 7].

### *Subjects and cell culture*

Overall, individual frequencies of breakage have been documented for sixty subjects. This sample includes 21 healthy subjects without exposure to environmental mutagens or carcinogens, 18 healthy subjects exposed to pesticides, 11 subjects exposed to radiation and free from any sign of disease, and 10 subjects exposed to radiation and affected by papillary thyroid carcinoma, not submitted to pharmacological treatments. For all subjects smoking, alcohol consumption and medications have been excluded; all were Caucasian. Informed consent was obtained from each subject prior to initiation of the study.

Peripheral blood cultures were established from all individuals using a total 0.3 ml of whole blood, added to 4.7 ml of Ham's F10 medium, supplemented with 1.5% phytohemoagglutinin, 10% fetal calf serum, and antibiotics. The cultures were incubated at 37°C for 72 hours; aphidicolin (final concentration 0.4 μM) was added for the last 26 hours of culture, according to the standard protocol for common fragile site expression. Colcemid (0.05 μg/ml) was added 90 minutes before cell harvesting and fixing. Chromosome preparations were set up according to standard protocols.

### *Co-expression analysis*

#### *Calculation of Spearman's correlation*

We evaluate the correlation between pairs of fragile sites (i.e. between the rows of the expression matrix defined above) using the non-parametric Spearman correlation coefficient. The choice of this type of correlation function (based on rank ordering) is due to the fact that in this way we do not need any assumption about the distribution followed by the entries of the expression matrix defined above. Another reason which led us to choose this rank-based type of correlation is that, due to its intrinsic robustness, it allows to eliminate possible bias in the data due to the inhomogeneity in the data sample. In this way we may neglect the fact that the data are obtained combining four different classes of subjects.

The Spearman rank-order correlation coefficient [31] is defined as follows:

$$r_s = \frac{\sum_i^N (R_i - \bar{R})(S_i - \bar{S})}{\sqrt{\sum_i^N (R_i - \bar{R})^2} \sqrt{\sum_i^N (S_i - \bar{S})^2}}$$

Here $N$ is the total number of subjects, $R_i$ is the rank of $x_i$ among the other $x$'s, $S_i$ is the rank of $y_i$ among the other $y$'s ($\bar{R}$ and $\bar{S}$ are the means of the ranks).

The significance of a nonzero value of $r_s$ is tested by computing

$$t = \sqrt{N-2} \frac{r_s}{\sqrt{1 - r_s^2}}$$

which is distributed approximately as Student's distribution with $N$ - 2 degrees of freedom. The comparison between the experimental and theoretical distributions is shown in Figure 4.

To sort out the co-expressed pairs of fragile sites that mainly cause such a difference between data and model we set several significance levels $\alpha$ = 1%, $\alpha$ = 5% and $\alpha$ = 10%. A correlation coefficient is evaluated as significant only if it satisfies to the constraint $t > t_{(1-\alpha)}$ where $t_{(1-\alpha)}$ is t at which the Student's cumulative density function respectively accounts for 99%($t_{(1-\alpha)}$ = 5.172), 95%($t_{(1-\alpha)}$ = 4.726) and 90%($t_{(1-\alpha)}$ = 4.529). In so doing, we perform a multiple hypothesis correction based on a naive Bonferroni test. This procedure let us operate a tight control on false positive occurrences. The Bonferroni method defines α' as a function of $M$, α' = α/M where $M = \frac{115*116}{2} = 6670$ is the number of correlation coefficients. This correction guarantees that the probability of having one or more false positives is no greater than α. Bonferroni corrected significance levels become respectively $\alpha' \approx 10^{-6}$, $\alpha' \approx 7 \cdot 10^{-6}$, $\alpha' \approx 10^{-5}$. Therefore, at the three significance levels, we retain only Spearman's correlation coefficients whose probability values are less than α'.

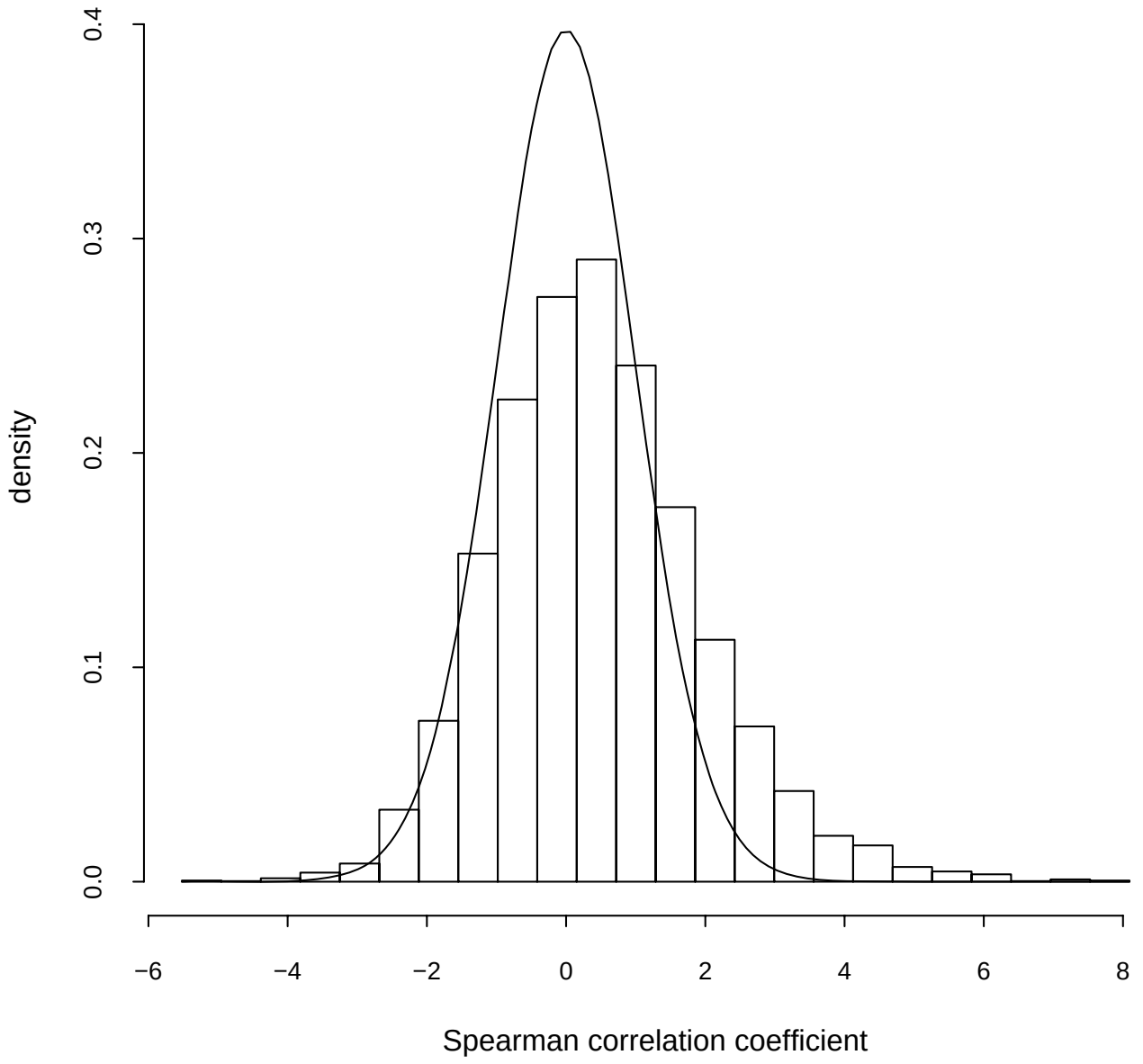

**Figure 4**
The theoretical t-Student distribution is superimposed on the histogram of t coefficients that we calculate from fragile site expression data.

A nice feature of our results is that they seem to be largely independent from the type of correlation function used to obtain them. We evaluated, as a test, the same correlations also using the Kendall function (another well known standard non-parametric test) finding essentially the same results.

These co-expression data can be usefully represented as a network in which nodes stand for fragile sites and links between couples of fragile sites are drawn if such fragile sites exhibit a significant Spearman's correlation coefficient in the sense we've just defined.

*Connected components reconstruction*

We extract connected components of the networks previously constructed by using the standard Hoshen-Kopelman algorithm [32]. This algorithm is one of the most efficient tools to find the connected components in an arbitrary undirected graph.

*Comparison with the random graph hypothesis*

The Erdos-Renyi random graph is the simplest possible model for a network. It depends on two parameters only: the number of vertices $n$ and the probability p of connecting two vertices with an edge. Actually this model describes not a single graph but an ensemble (in the sense of statistical mechanics) of graphs in which a graph with exactly n vertices and $m$ edges appears with probability $p^m (1 - p)^{M-m}$ where $M = \frac{n(n-1)}{2}$ is the number of pairs of vertices of the graph (and hence the maximum possible number of edges). The most important feature of the model is the presence at a particular value of p of a phase transition called percolation transition in which suddenly a giant connected component appears in the graph. This transition occurs exactly at z = 1 (where z is the mean degree of the graph and is given by $z = p(n-1)$).

It is easy to see that the three thresholds discussed in the text correspond to networks with a link densities $p$ = 0.0069, $p$ = 0.0094 and $p$ = 0.0126 respectively i.e. to a mean expected degree $z$ = 0.79, $z$ = 1.08 and $z$ = 1.45.

While for the lowest threshold the mean degree is in the percolating phase and so it is not surprising that we find a giant connected component in the graph, the mean degree for the highest threshold $z$ = 0.79 is far below the percolation threshold, thus the fact that also in this case a large connected component appears is a highly non trivial result.

Another important feature of random graphs is that, due to their simplicity, it's rather easy to evaluate a number of important graph theoretical quantities. In particular in our analysis we used the probability of a vertex having a degree $k$ $p_k = \binom{n}{k} p^k (1-p)^{n-k} \cong \frac{z^k e^{-z}}{k!}$ and the mean clustering coefficient which (for an undirected graph) is defined as:

$$<C> = \frac{\sum_{i=1}^{n} C_i}{n} \quad \text{where} \quad C_i = \frac{2|\{e_{jk}\}|}{K_i(K_i-1)}$$

where $e_{jk}$ denotes an edge between vertices $v_k$ and $v_j$ which are among the nearest neighbours of the vertex $v_i$ (degree $K_i$).

***Community structure of the network***

*Algorithm*

To display the community structure of the network we apply the agglomerative hierarchical clustering algorithm proposed by Newman [33]. The starting state has each vertex standing isolated. The joining of communities together in pairs is chosen so that it results in the greatest increase (or smallest decrease) in the modularity coefficient Q (see next section for the exact definition). The best partition of the network in communities corresponds to the maximal value of Q.

*Validation of the community structure*

The so called "modularity coefficient" is defined as $Q = \sum_i (e_{ii} - a_i^2)$ where $e_{ij}$ is the fraction of edges in the network that connect vertices of the community $i$ with those of the community $j$. Roughly speaking Q measures the fraction of edges which lie within the community minus the expected value for the same quantity in a random graph, thus for a random graph Q = 0 while larger values of Q indicate a significant departure from a random distribution of the edges. It is interesting to observe that for the connected components at the two highest thresholds Q is compatible with zero, thus indicating that no substructure is present in these components. On the contrary at the lowest threshold Q is definitely larger than 0 thus indicating that the large connected component that we find at this threshold is actually the combination of two separate subcomponents (which almost exactly coincide with those that we find at the highest thresholds).

#### *Functional characterization of connected components by Gene Ontology (GO) filter*

Cytogenetic definition of fragile sites is adopted except when a more precise identification of boundaries could be found in the literature [see Additional file 8]. The NCBI database for gene specific information offers positional information on all recognized common fragile sites. The first step is the construction of the sets of genes located at fragile sites belonging to the connected components that are detected at each significance level. We produce a mapping of the genes to their corresponding fragile sites by the means of the data mining Biomart tool provided by the Ensembl [34] database [41] (Ensemblv36).

We perform a bioinformatics analysis using the Gene Ontology functional annotation scheme [38], version 3.1191) to investigate potential correlations between the function, biological role and cellular location of the protein products of genes and their location at fragile sites. We filter the set of genes by each of the three main subontologies (*biological process*, *molecular function* and *cellular component*) separately. We always consider a gene annotated to a certain GO term and to all its ascendants in the GO hierarchy. For every GO term the number of associated genes within the set is calculated. We perform an exact Fisher's test to check whether the term appears in the set significantly more often than expected by chance. Indeed the Fisher's test gives the probability P of obtaining an equal or greater number of genes annotated to each term in a set made of the same number of genes but selected at random from the full list of annotated genes in the human genome. For a given GO term t let K(t) be the total number of genes annotated to it in the genome and k(t) the number of genes annotated to it in the set. If J and j denote the number of genes in the human genome and in the set respectively, such probability is given by the right tail of the hypergeometric distribution:

$$P(J,K(t),j,k(t)) = \sum_{h=k(t)}^{\min(j,K(t))} F(J,K(t),j,h)$$

where

$$F(J,K,j,h) = \frac{\binom{K}{h}\binom{J-K}{j-h}}{\binom{J}{j}}$$

If P is statistically significant, then we can postulate the existence of a correlation between the overrepresentation of the term and the functional characterization of the gene set.

Note that, as we perform the Fisher's test for above 19000 GO terms, adjustment of estimates of statistical significance for multiple testing are needed. In agreement with our previous experience [35,36] where we deal with this issue, we adopt $P = 10^{-4}$ as the threshold on the Fisher's test P-values.

##### *Genomic localization effects on Gene Ontology outcomes*

Chromosomal localization of genes annotated to statistically significant GO terms is worthy of caution because it might somehow bias the importance of GO outcomes at two levels.

First, the occurrence of a p-value above the threshold of acceptance may sometimes reflect the presence of clusters of functional families at a few fragile sites. Therefore the extension of this functional feature to the full connected component might be a forced interpretation of the actual outcomes. In our wok our decision to accept or not such cases depends on whether or not a GO term adds to a generalized functional characterization of the connected component. Let us mention two examples. Both '*response to virus*' and '*interferon-α/β receptor binding*' ought their acceptable p-values to the co-localization of several members of the interferon-α/β family at 9p21.3. However we report these GO terms because they agree with a number of other outcomes as described in Results and Discussion.

## Authors' contributions

AR: implementation of the algorithms, data analysis and participation in the project design.

DC: Gene-Ontology analysis of the data.

AP: participation in the discussion on biological significance of results.

MC: project design and supervision of the computational part.

IS: project design and supervision.

All the authors contributed to the manuscript writing, read and approved the final version.

## Additional material

### Additional file 1

*Gene Ontology characterization of the connected component $A_1$ at $\alpha$ = 1%. Gene Ontology characterization of the connected component $A_1$ when the significance level for fragile site correlation is set to 1%. Significantly over-represented GO words are associated the full set of annotated genes. Genes' identifiers provided by the Hugo Gene Nomenclature Committee (Hugo ids) and genes' localization in fragile sites are reported.*

Click here for file

[http://www.biomedcentral.com/content/supplementary/1471-2105-7-413-S1.pdf]

### Additional file 2

*Gene Ontology characterization of the connected component $B_1$ at $\alpha$ = 1% and $\alpha$ = 5%. Gene Ontology characterization of the connected component $B_1$ when the significance level for fragile site correlation is set to 1% and to 5%. Significantly over-represented GO words are associated the full set of annotated genes. Genes' identifiers provided by the Hugo Gene Nomenclature Committee and genes' localizations in fragile sites are reported.*

Click here for file

[http://www.biomedcentral.com/content/supplementary/1471-2105-7-413-S2.pdf]

### Additional file 3

*Gene Ontology characterization of the connected component $A_5$ at $\alpha$ = 5%. Gene Ontology characterization of the connected component $A_5$ when the significance level for fragile site correlation is set to 5%. Significantly over-represented GO words are associated the full set of annotated genes. Genes' identifiers provided by the Hugo Gene Nomenclature Committee and genes' localizations in fragile sites are reported.*

Click here for file

[http://www.biomedcentral.com/content/supplementary/1471-2105-7-413-S3.pdf]

### Additional file 4

*Gene Ontology characterization of the connected component $ABC_{10}$ at $\alpha$ = 10%. Gene Ontology characterization of the connected component $ABC_{10}$ when the significance level for fragile site correlation is set to 10%. Significantly over-represented GO words are associated the full set of annotated genes. Genes' identifiers provided by the Hugo Gene Nomenclature Committee and genes' localizations in fragile sites are reported.*

Click here for file

[http://www.biomedcentral.com/content/supplementary/1471-2105-7-413-S4.pdf]

### Additional file 5

*Gene Ontology characterization of the connected component $D_{10}$ at $\alpha$ = 10%. Gene Ontology characterization of the connected component $D_{10}$ when the significance level for fragile site correlation is set to 10%. Significantly over-represented GO words are associated the full set of annotated genes. Genes' identifiers provided by the Hugo Gene Nomenclature Committee and genes' localizations in fragile sites are reported.*

Click here for file

[http://www.biomedcentral.com/content/supplementary/1471-2105-7-413-S5.pdf]

### Additional file 6

*Full list of triangles detected at $\alpha$ = 1%. The full list of triangles detectable at the significance level for fragile site correlated expression set to $\alpha$ = 1%.*

Click here for file

[http://www.biomedcentral.com/content/supplementary/1471-2105-7-413-S6.pdf]

### Additional file 7

*Fragile sites expression matrix. Data on fragile sites expression are embeded in a matrix* M [116] [60] *whose* $m_{ij}$ *element represents the absolute number of breakage events that affect the fragile site* i *in the subject* j *(each entry of the matrix refers to 100 metaphases). The rows in the matrix correspond to the fragile sites and the columns to the different subjects.*

Click here for file

[http://www.biomedcentral.com/content/supplementary/1471-2105-7-413-S7.txt]

### Additional file 8

*Cytogenetic position of fragile sites. Cytogenetic position of fragile sites is reported in the first column; when cytogenetic regions where fragility has been observed include annotated fragile sites, we specify annotated fragile sites along with their genomic position. Fragile sites' genomic lengths and overall breakage occurrences are provided as well.*

Click here for file

[http://www.biomedcentral.com/content/supplementary/1471-2105-7-413-S8.pdf]

## Acknowledgements

The authors thank Leonardo Cocchi for expert technical assistance and Ferdinando Di Cunto, Paolo Provero and Tiziana Venesio for useful discussions. This work is partially supported by the Fund for Investments of Basic Research (FIRB) from the Italian Ministry of the University and Scientific Research, No. RBNE03B8KK-006.